%% file: TAPAS-3.tex
\documentclass{aa}  

\usepackage{graphicx}
\usepackage{txfonts}
\usepackage{tabu}
\usepackage{color}
\def\kms{\hbox{$\thinspace {\mathrm{km~s^{-1}}}$}}
\def\ms{\hbox{$\thinspace {\mathrm{m~s^{-1}}}$}}
\def\ALi{\hbox{$\thinspace A (\mathrm{Li})$}}

\def\shk{\hbox{$\thinspace S_{\mathrm{HK}}^{\mathrm{inst}}$}}
\def\MJ{\hbox{$\thinspace M_{\mathrm{J}}$}}

\def\ms{\hbox{$\thinspace {\mathrm{m~s^{-1}}}$}}
\def\kms{\hbox{$\thinspace {\mathrm{km~s^{-1}}}$}}

\begin{document}

  \title{Tracking Advanced Planetary Systems (TAPAS) with HARPS-N. 
  \thanks{Based on observations obtained with the Hobby-Eberly Telescope,
 which is a joint project of the University of Texas at Austin, the Pennsylvania State University, 
Stanford University, Ludwig-Maximilians-Universit\"at M\"unchen, and Georg-August-Universit\"at G\"ottingen.}
\thanks{Based on observations made with the Italian Telescopio Nazionale Galileo (TNG) operated 
on the island of La Palma by the Fundaci\'on Galileo Galilei of the INAF (Istituto Nazionale di Astrofisica) 
at the Spanish Observatorio del Roque de los Muchachos of the Instituto de Astrof\'{\i}sica de Canarias.}
}
   \subtitle{III. HD 5583 and BD+15 2375 - two cool giants with warm companions.}
   \titlerunning{TAPAS. Two cool giants with warm companions.}
   \authorrunning{A. Niedzielski et al.}

   \author{A. Niedzielski
          \inst{1}  
                              \and
          E. Villaver
         \inst{2}
                \and
         G. Nowak
          \inst{3,4,1} 
                                      \and
         M. Adam\'ow
          \inst{5,1}
                                                \and  
         K. Kowalik
          \inst{6}        
                                       \and                         
          A. Wolszczan
          \inst{7,8}
                             \and
         B. Deka-Szymankiewicz
          \inst{1}
                                      \and
        M. Adamczyk
          \inst{1}
                                                        \and              
         G. Maciejewski
          \inst{1}
                   }
                   
   \institute{Toru\'n Centre for Astronomy, Faculty of Physics, Astronomy and Applied Informatics, Nicolaus Copernicus University in Toru\'n, Grudziadzka 5, 87-100 Toru\'n, Poland.
              \email{Andrzej.Niedzielski@umk.pl}
                                    \and
        Departamento de F\'{\i}sica Te\'orica, Universidad Aut\'onoma de Madrid, Cantoblanco 28049 Madrid, Spain.
         \email{Eva.Villaver@uam.es}
             \and
              Instituto de Astrof\'isica de Canarias, E-38205 La Laguna, Tenerife, Spain.
              \and
              Departamento de Astrof\'isica, Universidad de La Laguna, E-38206 La Laguna, Tenerife, Spain.            
         \and
McDonald Observatory and Department of Astronomy, University of Texas at Austin, 2515 Speedway, Stop C1402, Austin, Texas, 78712-1206, USA.
\and
              \and
National Center for Supercomputing Applications, University of Illinois, Urbana-Champaign, 1205 W Clark St, MC-257, Urbana, IL 61801, USA   
             Department of Astronomy and Astrophysics, Pennsylvania State University, 525 Davey Laboratory, University Park, PA 16802, USA.
          \email{alex@astro.psu.edu}
         \and
             Center for Exoplanets and Habitable Worlds, Pennsylvania State University, 525 Davey Laboratory, University Park, PA 16802, USA.
             }

   \date{Received;accepted}

 
  \abstract
   {Evolved stars are crucial pieces  to understand the dependency 
   of the planet formation mechanism on the stellar mass and to explore deeper the mechanism involved in star-planet interactions.
    Over the past ten years, we have monitored about 1000 evolved stars for 
   radial velocity variations in search for low-mass companions under the Penn State - Torun 
   Centre for Astronomy Planet Search program with the Hobby-Eberly Telescope. 
   Selected prospective candidates that required higher RV precision measurements 
   have been followed with HARPS-N at the 3.6 m Telescopio Nazionale Galileo 
   under the TAPAS project. }
   {We aim to detect planetary systems around evolved stars to be able to build sound 
   statistics on the frequency and intrinsic nature of these systems, and to deliver in-depth 
   studies of selected planetary systems with evidence of star-planet interaction processes. }
   {For HD 5583 we obtained 14 epochs of precise RV measurements 
   collected over 2313 days with the Hobby-Eberly Telescope (HET), and 22  epochs 
   of ultra-precise HARPS-N data collected over 976  days. For BD+15 2375 we collected 24 epochs of HET data over 3286 days and 25 epochs of HARPS-S data over 902 days.}
   {We report the discovery of two planetary mass objects orbiting two evolved Red Giant stars: 
   HD~5583 has a $m \sin i = 5.78 \MJ$ companion at 0.529~AU in a nearly circular orbit ($e=0.076$), the closest companion to a giant star detected with the RV technique,
   and BD+15~2735 that with a m sin i= 1.06 Mj holds the record of the lightest planet found so far orbiting an evolved star 
    (in a circular $e=0.001$, 0.576~AU orbit). These are 
   the third and fourth planets found within the  TAPAS project, a HARPS-N monitoring of evolved 
   planetary systems identified with the Hobby-Eberly Telescope.}
   {}

   \keywords{Stars: evolution,  activity, low-mass, late-type, planetary systems; Planets and satellites: individual: HD 5583 b, DB+15 2735 b.  }

   \maketitle
%

\section{Introduction}

The discovery of the first exoplanet orbiting a solar-type star  \citep{Mayor1995} unveiled a completely 
unexpected population of massive planets orbiting within 1 AU of their stars (the so-called  hot Jupiters). 
{Currently, we know that about 0.5\% of stars hosts such objects \citep{Wright2012} and the class is limited
to gas planets within 0.1~AU or, roughly, with orbital periods of up to 10 days}. The origin of those planets 
is still debated, as  prevalent planet formation theories 
\citep[e.g.][]{Pollack1996,Boss1997} do not allow for {\it in situ} formation of such objects.
Two possible mechanisms have been proposed to explain the orbital distribution of hot Jupiters: i) 
early proto-planetary disc migration \citep{Lin1996}; and ii) dynamical interaction with a third massive 
object in a multiple system that leads to a smaller, eccentric, orbit that circularizes through tidal
dissipation \citep{Rasio1996a}. Interesting is the recent estimate by \cite{Ngo2015} of the large fraction
of  hot Jupiters ($72\%\pm16\%$) being part of multi-planet and/or multi-star systems.  
An intriguing class of objects in this context are the so-called warm-Jupiters, gas planets 
at 0.1 to 0.5 AU, with orbital periods of 10 to 100 days, as transition objects possibly
feeding the hot-Jupiter population \citep[e.g.][]{Howard2012}. 

The growing statistics of hot and of warm Jupiters mainly comes  from transit planet searches that deliver 
the population of low-mass companions to Main Sequence stars. In the case of more evolved stars, like subgiants 
and giants, the fact that close companions are very rare was already noted by 
by \cite{Johnson2007, 2007ApJ...660..845B, 2008PASJ...60..539S}. As a matter of fact, there is no hot
Jupiter known around a giant and only very few warm Jupiters orbit subgiants \citep[e.g.][]{Lillo2014,Johnson2010}.  
One possible explanation of this fact is that low-mass close companions to evolved stars are ingested by their hosts 
due to tidal interactions \citep{2008PASJ...60..539S,VillaverLivio2009}. There is, however,
an alternative explanation and it is related to the short depletion timescales of gas in the proto-planetary
disks expected in massive stars  \citep{2007ApJ...660..845B,Currie2009}. We would like to bring attention 
here to the fact that evolved stars are not necessarily intermediate-mass stars and these terms should
not be used interchangeably. Furthermore, early gas-disk depletion will only work as an explanation for 
the lack of close, low mass, companions to evolved stars if the bulk of the population of hot Jupiters 
builds via early proto-planetary disk migration and all evolved stars have massive progenitors.

An additional complication to the observational interpretation of the data comes from the fact that in the case 
of the evolved giants past the RGB-tip, the planetary systems may not only be affected by previous 
engulfment process \citep{VillaverLivio2007, Kunitomo2011} but also, in some cases, experience 
RGB-tip planetary orbit shrinking  \citep{VillaverLivio2009,Villaver2014} and possibly rebuilding that 
way the warm-Jupiter population. Furthermore, stellar mass-loss can trigger delayed
instability in multiple planetary systems \citep{Mustill2014}.

Statistical considerations including both short and long period systems around those stars are fundamental to determine the actual process responsible for the lack of hot Jupiters 
around evolved stars 
since we  cannot exclude that the observed distribution might be due to a selection resulting from degradation of effective RV precision by stellar activity \citep{Niedzielski2015c}.
Building meaninful statistics of those objects is the aim of several observing projects 
like the McDonald Observatory Planet Search \citep{CochranHatzes1993, HatzesCochran1993}, 
the Okayama Planet Search \citep{Sato2003}, 
the Tautenberg Planet Search \citep{2005A&A...437..743H},
the Lick K-giant Survey \citep{2002ApJ...576..478F},  
the ESO Ferros planet search \citep{SetiawanHatzes2003, SetiawanPasquini2003}, 
Retired A Stars and Their Companions \citep{Johnson2007},
the Coralie $\&$ HARPS search \citep{2007A&A...472..657L},
the Boyunsen Planet Search \citep{Lee2011}, our own 
Pennsylvania-Toru\'n Planet Search  (PTPS, \citealt{Niedzielski2007, NiedzielskiWolszczan2008}),
and several others. 

Properties of planetary systems around evolved hosts have been explored in several papers already.
The planet-metallicity correlation seems to be well established for the massive evolved stars
\citep{Maldonado2013,Reffert2015, Jofre2015} and doubtful for the lower mass stars \citep{Maldonado2013}
or planet candidates samples  \citep{Reffert2015}. And while several studies so far have 
expected differences in the orbital distribution of planets for massive stars mostly due to
a faster depletion of the gas in the proto-planetary disk (see e.g. 
\citealt{2009ApJ...695.1210K, 2007ApJ...660..845B, Currie2009, 2015arXiv150200631R}). 
It is very difficult to observationally disentangle the primordial differences from the evolutionary
ones that should also leave an imprint in the orbital distribution of the evolved systems (see e.g. \citealt{Villaver2014}).

The relation between the planetary properties and those of the central stars offer important clues
to our current understanding of planet formation and guide the theoretical modeling.
In the case of giant stars the discovery of new planetary mass objects is of special relevance 
for several reasons. New objects help to populate the numbers needed to build better statistics 
of planets around evolved stars. Many of these planets orbit stars with masses $M > 1.2-1.5 M_\odot$ 
and since a clear dependency between planet formation 
and the stellar mass has not been found yet, any new planet found around a massive 
star adds valuable information to our understanding. In this regard, knowing the evolutionary status 
of the star is fundamental, given that the primordial properties of these planet-star systems 
can be somehow erased as the star leaves the main sequence \citep{VillaverLivio2007, VillaverLivio2009}. 
Furthermore, evolved red giants require of precise evolution history studies for 
a proper interpretation of the observed properties of their planetary systems. 

In this paper, 
third of the series (after \cite{Niedzielski2015a}, hereafter Paper~I, and \citealt{Adamow2015}) 
we present two more close-orbit companions to red giants from our HARPS-N follow-up of selected PTPS targets under the TAPAS project. 
We discuss their future in terms 
of how stellar evolution is expected to change their orbital properties, and explore their past 
by contextualizing these two systems into and the general population of warm Jupiter planets 
observed around main sequence stars.

The paper is organized as follows: in
Sect. \ref{observations} we present the observations obtained for
our  targets and outline the reduction and
measurement procedures; Sect. \ref{results-g} shows the results of
the Keplerian data modeling; in Sect. \ref{activity} we extensively discuss the influence of the stellar activity
on the RV variation measurements; and in Sect.
 \ref{conclusions}, we discuss the results of our
analysis and present the conclusions of this work.


\section{Targets, observations and the RV data\label{observations}}

The spectroscopic observations presented in this paper were made  with
the 9.2  Hobby-Eberly Telescope
(HET, \citealt{Ramsey1998}) and its  High-Resolution Spectrograph (HRS,
\citealt{Tull1998}) in the queue scheduled mode \citep{Shetrone2007},
and with the 3.58 meter Telescopio Nazionale Galileo (TNG) and its High
Accuracy Radial velocity Planet Searcher in the North hemisphere (HARPS-N,
\citealt{Cosentino2012}). 

Both HD~5583 and BD+15~2375 belong to a sample of about 300 planetary or brown dwarf (BD) candidates 
identified in the complete sample of over 1000 stars observed with HET 
and HRS  since 2004 within {PennState - Toru\'n Centre
for Astronomy Planet Search} (PTPS, \citealt{Niedzielski2007, NiedzielskiWolszczan2008, Niedzielski2015b}) 
and selected for a more intense precise RV follow-up. 
The program {Tracking
Advanced Planetary Systems (TAPAS) with HARPS-N}  is the result of intensifying
the monitoring of a selected number of about 100 PTPS-identified targets, that is, 
those with potentially  multiple and/or low-mass companions, with
expected p-mode oscillations of a few $\ms$, and
systems with  evidence of recent or future {star-planet} interactions
(Li-rich giants, low-orbit companions, etc.). 

{ HD~5583 and  BD+15~2375 are giant stars selected as a TAPAS targets 
because the available HET observations pointed to  relatively short orbital periods,
making these stars promising targets
for star-planet interactions studies.}

\subsection{Observations}

The HET and HRS spectra were gathered with the HRS fed
with a 2 arcsec fiber, working in the R=60\,000 mode with a gas cell
($I_2$) inserted into the optical path.  
The configuration and observing procedure employed in our program were, in practice, 
identical to those described by \cite{Cochran2004}. Details of our survey, the observing procedure,
and data analysis have been described in detail elsewhere \citep{Niedzielski2007, Nowak2013, Niedzielski2015b}.  
We use a combined gas-cell  \citep{MarcyButler1992, Butler1996}, and  cross-correlation
\citep{Queloz1995, Pepe2002} method for precise RV and Spectral Line Bisector (BS) measurements, respectively.  
The implementation of this technique to our data is described in \cite{Nowak2012} and \cite{ Nowak2013}. 

The HARPS-N, a near-twin of the HARPS instrument mounted
at the ESO 3.6~m telescope in La Silla \citep{Mayor2003},  is an echelle spectrograph covering the visible
wavelength range between 383~nm and 693~nm with a resolving power of $\mathrm{R}\sim115\;000$.  
Radial velocity measurements and their uncertainties, as well as BS, were obtained
with the standard user pipeline, which is based on the weighted
CCF method \citep{1955AcOpt...2....9F, 1967ApJ...148..465G, 1979VA.....23..279B, Queloz1995, Baranne1996, 
  Pepe2002}. To obtain highest precision RV, we used the simultaneous Th-Ar calibration 
mode of the spectrograph.  RVs were obtained with the K5 
cross-correlation mask. 

More details on the instrumental configuration were presented in Paper~I.

\begin{table}
\centering
\caption{Summary of the available data on HD 5583.}
\begin{tabular}{lll}
\hline
Parameter & value & reference\\
\hline
\hline
$V$  [mag]& 7.60$\pm$0.01 &  \cite{Hog2000} \\
$B-V$ [mag] & 0.94 $\pm$ 0.03 & \cite{Hog2000} \\ 
$(B-V)_0$ [mag] & 0.93& \cite{Niedzielski2015c}\\
$M_\mathrm{V}$ [mag] & 0.65& \cite{Niedzielski2015c} \\
 $T_{\mathrm{eff}}$ [K] & 4830$\pm$45 & \cite{Niedzielski2015c} \\
 $\log g$ & 2.53$\pm$0.14& \cite{Niedzielski2015c} \\
$[Fe/H]$ & -0.50$\pm$0.18 & \cite{Niedzielski2015c}\\
RV $[\kms]$ & 11.68 $\pm$ 0.03 & \cite{Nowak2012} \\
$v_{\mathrm{rot}} \sin i_{\star}$ $[\kms]$ & 2.2$\pm$2.3 & \cite{Adamow2014} \\
$\ALi $& $<0.22$ &  \cite{Adamow2014} \\
$[$O/H$]$ & -0.03$\pm$0.13 & \cite{Adamow2014phd}\\
$[$C/H$]$ & -0.58$\pm$0.04 & this work\\
$[$Ba/H$]$ & -0.65                & this work\\
$[$La/H$]$ & -0.53$\pm$0.07 & this work\\
$[$Eu/H$]$ & -0.16 & this work\\
\hline
$M/M_{\odot}$ & 1.01$\pm$0.1 & \cite{Niedzielski2015c}\\
$\log L/L_{\odot}$ & 1.61$\pm$0.09 & \cite{Niedzielski2015c}\\
$R/R_{\odot}$ & 9.09$\pm$1.5 & \cite{Niedzielski2015c}\\
$\log \mathrm{age}$ [yr]& 9.87$\pm$0.15& \cite{Niedzielski2015c}\\
$d$ [pc] &  221 $\pm$ 14 & calculated from M$_{V}$\\
$V_{\mathrm{osc}}$ [$\ms$] & 9.4$^{+4.2}_{-2.9}$ & this work\\
$P_{\mathrm{osc}}$ [d] & 0.28$^{+0.15}_{-0.11}$ & this work\\
$P_{\mathrm{rot}}$ [d] & 209$\pm$221 & this work\\ 
\hline
\hline
\end{tabular}
\label{Parameters1}
\end{table}


\begin{table}
\centering
\caption{Summary of the available data on BD+15 2375.}
\begin{tabular}{lll}
\hline
Parameter & value & reference\\
\hline
$V$  [mag] & 10.31$\pm$0.023 &Hagkvist $\&$ Oja (1973)\\
$B-V$ [mag] & 1.05$\pm$0.008 &Hagkvist $\&$ Oja (1973)\\
$(B-V)_0$ [mag] & 1.06 & \cite{Zielinski2012} \\
$M_\mathrm{V}$ [mag] & 0.80 & \cite{Zielinski2012} \\
 $T_{\mathrm{eff}}$ [K] & 4649$\pm$30 & \cite{Zielinski2012} \\
 $\log g$ & 2.61$\pm$0.12& \cite{Zielinski2012} \\
$[$Fe/H$]$ & -0.22$\pm$0.08 & \cite{Zielinski2012}\\
RV $[\kms]$ & -9.646$\pm$0.028& \cite{Zielinski2012} \\
$v_{\mathrm{rot}} \sin i_{\star}$ $[\kms]$& 2.5$\pm$0.5 &\cite{Adamow2014}\\
$\ALi $& $<$0.55 &  \cite{Adamow2014} \\
$[$O/H$]$ & 0.11$\pm$0.20 &\cite{Adamow2014} \\
$[$C/H$]$ & -0.23$\pm$0.06 & this work\\
$[$Ba/H$]$ & -0.24 & this work\\
$[$La/H$]$ & -0.15$\pm$0.02 & this work\\
$[$Eu/H$]$ & 0.07 & this work\\
\hline
$M/M_{\odot}$ & 1.08$\pm$0.14 & \cite{Adamczyk2015}\\
$\log L/L_{\odot}$ & 1.57$\pm$0.10 & \cite{Adamczyk2015}\\
$R/R_{\odot}$ & 8.95$\pm$1.45& \cite{Adamczyk2015}\\
$\log \mathrm{age}$  [yr]& 9.87$\pm$0.18& \cite{Adamczyk2015}\\
$d$ [pc] &  774 $\pm$ 165 & calculated from M$_{V}$\\
$V_{\mathrm{osc}}$ $[\ms]$ & 8.1$^{+4.3}_{-2.7}$ & this work\\
$P_{\mathrm{osc}}$ [d] & 0.25$^{+0.14}_{-0.10}$ & this work\\
$P_{\mathrm{rot}}$ [d] & 181$\pm$ 47 & this work\\

\hline

\end{tabular}
\label{Parameters2}
\end{table}

\subsection{Stars and their parameters}

HD 5583 (BD+34~154, 
TYC~2285-00548-1) is a $V=7.60\pm 0.01$ 
and $B=8.54\pm0.02$ \citep{Hog2000} 
star in Andromeda with no parallax available. It belongs to the PTPS ''Giants and Subgiants'' sample 
and was included in PTPS since May 2006. Its atmospheric parameters were determined 
with the purely spectroscopic method  of \citet{Takeda2005a, Takeda2005b} by \cite{Niedzielski2015c}.

BD+15 2375 (GSC 00870-00204, 
TYC 0870-00204-1) is a $V =10.31\pm 0.023$ 
and $B-V =1.05\pm0.008$ \citep{1973A&AS...12..381H}  star in Leo with no parallax available.
The star belongs to PTPS ''Red Giant Clump" sample and was observed within PTPS since March 2004.
Its atmospheric parameters were estimated means of the method of \citet{Takeda2005a, Takeda2005b}  by \cite{Zielinski2012}. 

Based on abundance calculations through the Spectroscopy Made Easy package \citep{SME1996} modeling of 27 spectral lines for  six
elements, \citep{Adamow2014, Adamow2014phd} obtained $v_{\mathrm{rot}} \sin i_{\star}$ 
and found no anomalies in the lithium abundances for none stars of the two stars presented in this paper when compared to the full sample of PTPS stars.

Stellar mass, luminosity, age and radii for HD~5583 and BD+15 2375~were estimated 
on the basis of spectroscopically determined atmospheric parameters 
in \cite{Niedzielski2015c} and \cite{Adamczyk2015} , respectively.

{ The amplitude and period of p-mode oscillations ($V_{\mathrm{osc}}$, $P_{osc}$) were estimated from the scaling relation by \cite{KjeldsenBedding1995}.}

{ With the rotational velocities from \cite{Adamow2014}  and stellar radii derived in  \cite{Adamczyk2015} we have obtained estimates of maximum rotation periods, $P_{rot}$.}

A summary of all the available data for HD~5583 and BD+15~2735  
is given in Tables \ref{Parameters1} and \ref{Parameters2}, respectively.

%
%
\begin{table}
\centering
\caption{HET and HRS RV and BS measurements  [$\ms$] of HD 5583}
\begin{tabular}{lrrrr}
\hline
MJD& RV  & $\sigma_{\mathrm{RV}}$ & BS  & $\sigma_{\mathrm{BS}}$ \\
\hline\hline 
54013.177743  &   -195.63    &    7.30   &    -7.81    &   13.07    \\
 54284.434317  &   -323.54   &     8.74   &    18.21  &     14.65   \\
 54336.287477 &     109.94    &    7.93   &   -43.09  &     23.75   \\
54443.218657   &  -131.48     &   8.20    &   -7.36   &    28.26   \\
54463.182975   &    98.26      &  5.11    &  -16.65   &    14.35    \\
55503.329329   &  -117.27     &   5.75   &    15.31    &   15.79   \\
55519.288877   &  -322.71     &   8.55   &   -40.63  &     18.67   \\
55856.354497   &   152.81    &    7.28   &    16.59   &    27.42   \\
55860.342268   &   191.88     &   5.28   &   -49.87   &    14.24   \\
55893.268194   &   104.24     &   6.73   &    28.19   &    18.70   \\
56231.335151   &  -178.57     &   5.74   &    50.54   &    14.26   \\
56289.183785   &   200.21      &  6.69   &   -35.89  &     21.92   \\
56310.108096   &    63.99    &    6.28   &    44.29   &    21.53   \\
56326.084780   &   -29.81   &    8.20  &    -15.01   &    25.38   \\
\hline
\end{tabular}
\label{HETdata1}
\end{table}
%
%
%

%
%
\begin{table}
\centering
\caption{TNG and HARPS-N RV and BS measurements [$\kms$] of HD 5583}
\begin{tabular}{lrrr}
\hline
MJD& RV &$\sigma_{\mathrm{RV}}$& BS  \\
\hline\hline

56261.9146166    &    12.13639 &    0.00129   &   0.09396        \\
56276.987077     &    12.13043  &   0.00142   &   0.07093         \\
 56293.9627115   &     12.28187 &   0.00086  &    0.08545        \\
56320.9140125    &    12.10671  &  0.00097 &    0.06084        \\
56502.2245484   &      11.89863  &   0.00161   &   0.05394        \\
56560.1937667   &      12.17773   &  0.00138  &    0.06409        \\
56647.0192778  &      11.80500  &   0.00222   &   0.07423        \\
56684.9002029  &      12.12080  &   0.00334  &    0.06706        \\
56895.1744165   &     11.92651  &   0.00167   &   0.09017        \\
56926.9761639   &     11.84783  &   0.00182  &    0.05790        \\
56927.2227979  &      11.86547  &   0.00184  &    0.06911        \\
56969.8459431  &      12.13113  &    0.00103  &    0.06408        \\
56970.054993    &    12.15012  &   0.00208   &   0.06180         \\
56970.0617981  &       12.15188   &  0.00171   &   0.06755        \\
56992.0106316  &      12.26357  &   0.00204  &    0.06316        \\
57034.813641    &    11.88753  &   0.00231    &  0.06620         \\
57034.9296223  &      11.87126  &   0.00287  &     0.06517        \\
57065.8378883   &     11.79888   &  0.00195   &   0.08513    \\    
57065.8635489    &    11.79112   &   0.00134   &   0.09514  \\    
     57196.2183147   &     11.78237   &     0.00137  &    0.07080 \\ 
     57238.1404346    &    12.02228    &    0.00127     &         0.07107  \\
     57238.2351083    &    12.02891   &    0.00135    &           0.06803  \\

\hline
\end{tabular}
\label{HARPSdata1}
\end{table}
%
%
%

%
%
\begin{table}
\centering
\caption{HET and HRS RV and BS measurements  [\ms] of BD+15 2375}
\begin{tabular}{lrrrr}
\hline
MJD& RV  & $\sigma_{\mathrm{RV}}$ & BS  & $\sigma_{\mathrm{BS}}$ \\
\hline\hline 
53093.348333   &   21.34   &     6.75    &   -7.785     &  16.56   \\
54198.322703   &   -30.45   &     7.06   &   -11.70   &    17.12   \\
54217.273848   &   -37.18   &     6.87   &    11.71   &    17.13   \\
54249.193831   &   -11.30   &     6.46   &    -1.93    &   14.63   \\
54564.141238   &    22.01  &      8.76   &    56.87  &     24.26   \\
54843.364103    &   -5.87   &    8.90   &   -17.09    &    9.50    \\
54874.296927    &   50.00   &     8.04   &    -5.54   &    21.10   \\
55173.461551   &     2.95    &    7.38   &   -26.85   &    19.26   \\
55232.313623   &     3.93   &     5.84   &    14.61   &    16.68   \\
55523.503021   &     2.65   &     8.39   &    -2.10    &   20.69    \\
55626.237488   &   -41.11   &     7.88  &    -41.63  &     19.31   \\
55648.351505   &     4.62    &    6.73   &   -20.52    &   17.85   \\
55696.221759   &    38.10   &     7.07   &    43.14   &    14.45   \\
55895.484062   &   -15.33   &     7.22   &   -32.80  &     14.97   \\
55929.390174   &   -36.91   &     6.96   &   -24.18  &     16.57    \\
55971.301088   &    31.42   &     6.30   &     9.35   &    16.01   \\
56075.190995   &     3.75    &    6.92   &    -3.29     &  15.60   \\
56081.172604   &    22.33  &      6.77  &     37.54   &    15.37   \\
56085.163505   &    27.29 &      6.94  &     24.96  &     15.85   \\
56319.523113   &    -8.43   &     7.82   &     5.45   &    21.94   \\
56331.303171   &   -42.86   &     7.32  &    -30.85   &    21.57   \\
56379.359236  &     -5.73    &    6.77   &     7.98   &    16.16   \\

\hline
\end{tabular}
\label{HETdata2}
\end{table}
%
%
%

%
%
\begin{table}
\centering
\caption{TNG and HARPS-N RV and BS measurements [\kms] of BD+15 2375}
\begin{tabular}{lrrr}
\hline
MJD& RV & $\sigma_{RV}$ & BS  \\
\hline\hline

56294.1928145    &   -9.28232 &    0.00109   &   0.09414        \\
56321.1212361    &   -9.33294 &     0.00113    &   0.09704        \\
56373.9922013    &   -9.32543  &   0.00113   &   0.07925        \\
56410.9474761    &   -9.31045  &   0.00114   &   0.09045        \\
56430.9456148   &    -9.28979  &   0.00161  &    0.09212        \\
56469.8872767   &    -9.31259  &    0.00232   &   0.09129        \\
56647.2299651   &    -9.34178  &   0.00257  &    0.08796        \\
 56685.164531    &    -9.30657  &   0.00303    &  0.08647         \\
56740.0370018   &    -9.27847  &   0.00259  &    0.08559        \\
56770.0052688   &    -9.31436  &   0.00203  &    0.09867        \\
56770.0431858   &    -9.30688  &    0.00181   &   0.09181        \\
56794.9549192   &    -9.35649  &   0.00241  &    0.09525        \\
56836.8943481    &   -9.33541  &   0.00300  &    0.07446        \\
57035.1756853   &    -9.28354  &   0.00224   &   0.09836        \\
57035.2550753   &    -9.30066  &   0.00268  &     0.10822        \\
57066.0277781   &    -9.28096 &   0.00334    &  0.09549 \\        
57066.0942389    &    -9.28313   &   0.00227  &    0.09654 \\      
57066.1775331    &   -9.30931   &  0.00334    &   0.09168 \\        
57066.2677252    &   -9.31988   &   0.00293  &    0.09983 \\       
57110.9410484   &    -9.35497    &    0.00256      &          0.07713 \\
57111.0951775    &   -9.31690     &   0.00821       &         0.06780 \\
57134.9450871   &    -9.38975    &    0.00162      &         0.09403 \\
57134.9972007   &    -9.36492    &    0.00116      &        0.09409 \\
57167.9442205    &   -9.29965     &   0.00136       &        0.09534 \\
57195.9206147    &   -9.27754     &   0.00149       &         0.09521 \\

\hline
\end{tabular}
\label{HARPSdata2}
\end{table}
\begin{figure}
   \centering
   \includegraphics[width=0.5\textwidth]{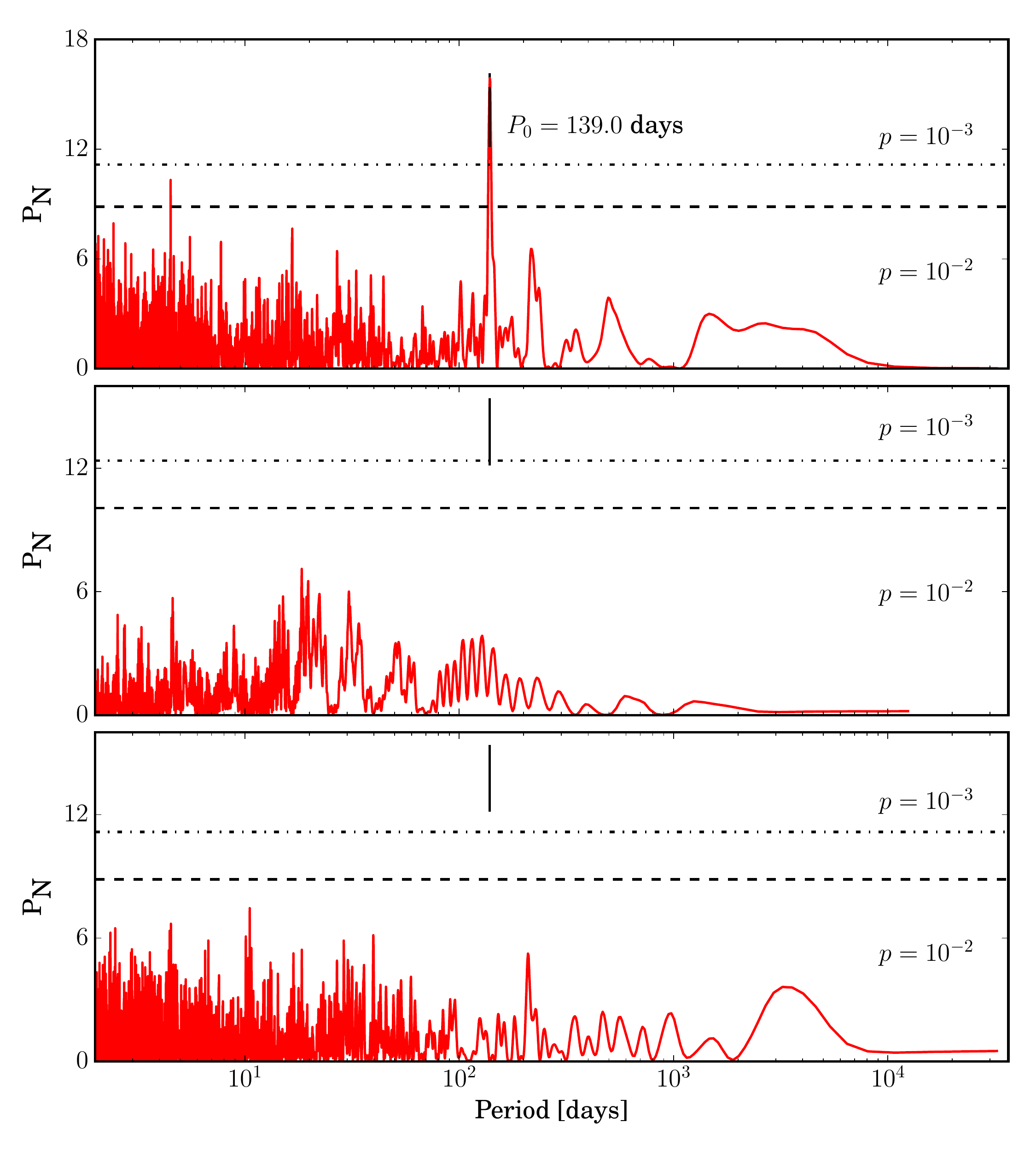}
   \caption{Lomb Scargle priodograms for HD 5583, top to bottom: combined RV data, 
   Super WASP photometry and RV residua. {\bf False alarm probability levels of p=0.01 and 0.001 are marked with horizontal lines.}}
   \label{LSP1}
\end{figure} 

\begin{figure}
   \centering
   \includegraphics[width=0.5\textwidth]{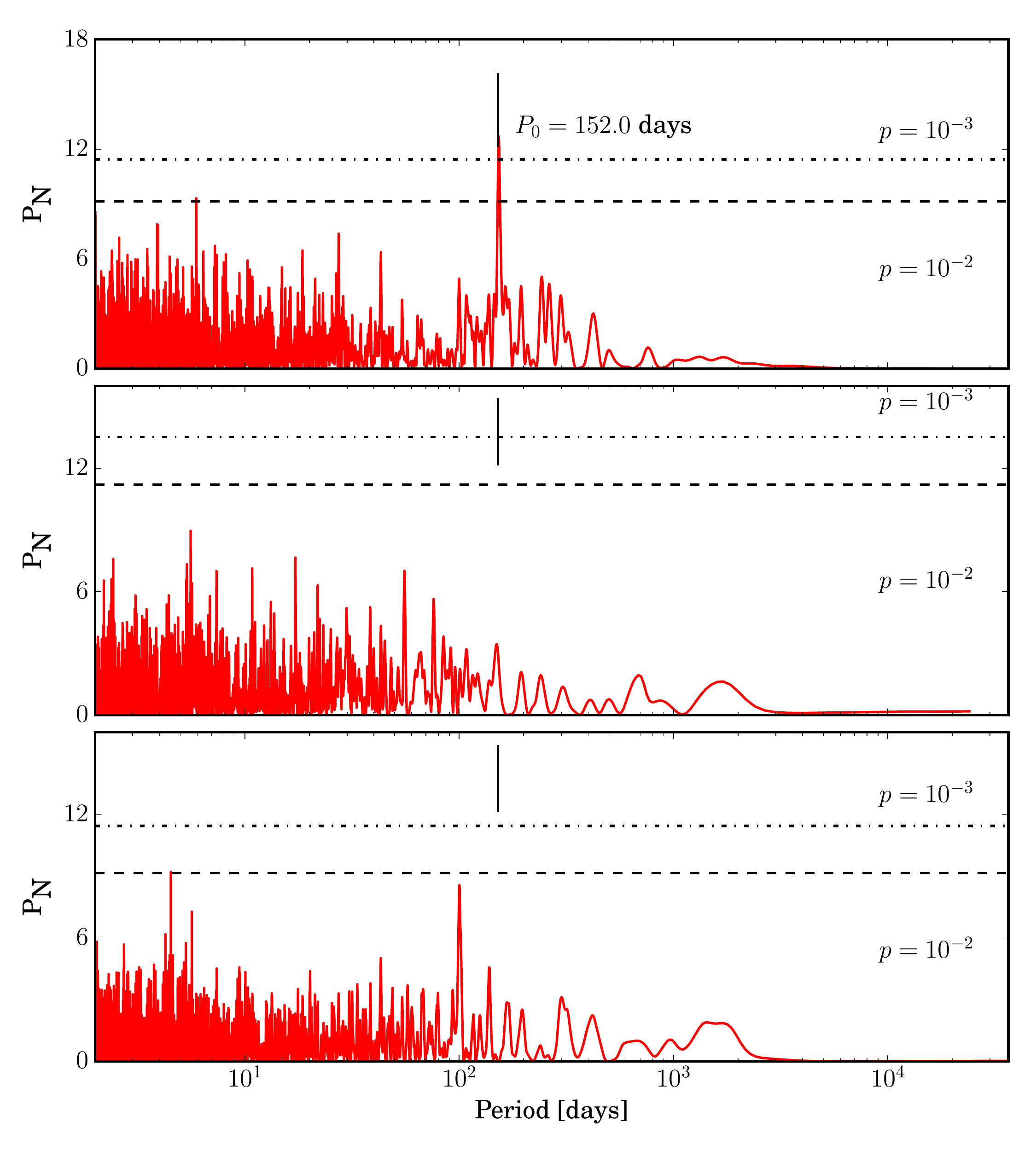}
   \caption{Lomb Scargle priodograms for BD+15~2735, top to bottom: combined RV data, ASAS photometry and RV residua. {\bf False alarm probability levels of p=0.01 and 0.001 are marked with horizontal lines.}}
   \label{LSP2}
\end{figure} 

\subsection{ Radial velocities  and spectral line bisectors}

HD 5583 was observed over 3225 days between Modified Julian Day MJD=54013 and 57238. 
We obtained 14 epochs of HET HRS RV and BS for HD~5583 over 2313 days between
MJD=54013 and 56326 and 22 epochs of TNG HARPS-N RV over 976 days between MJD=56261 and 57238.

The HET RV data (Table \ref{HETdata1}) show a peak-to-peak amplitude of $524 \ms$ 
and average uncertainty of $7 \ms$ while the BS varies by $100 \ms$ with average
uncertainty of $19 \ms$. The Parsons correlation coefficient between RV and BS is -0.21. 
The obvious lack of correlations justifies the interpretation of RV variations as due to Keplerian motion. 

HARPS-N RV for this star (Table \ref{HARPSdata1}) show an amplitude of $499 \ms$
and average uncertainty of $1.7 \ms$. The BS varies by $41 \ms$. The correlations 
coefficient between RV and BS  of only -0.13 justifies again the  interpretation
of the observed RV variations as Keplerian motion.

We observed BD+15 2375 over 4103 days between MJD=53093 and 57196. 
We gathered 24 epochs of HET HRS RV data over 3286 days  between MJD=53093 and 56379 
of which two, apparent outliers were rejected. In addition we collected 25 epochs 
of TNG HARPS-N RV over 902 days between 56294 and 57196.

The available HET HRS RV data (Table \ref{HETdata2}) show an amplitude of $93 \ms$ 
and average uncertainty of $7 \ms$, while the BS varies by $99 \ms$ with average uncertainty
of $17 \ms$. The correlation coefficient between RV and BS is 0.60, just above the critical
value of 0.54, so the relation between RV and BS requires more detailed examination (see Section \ref{activity}).

HARPS-N RV collected for this star (Table \ref{HARPSdata2}) show an amplitude of $112 \ms$
and average uncertainty of $2.4 \ms$. The BS varies by $40 \ms$ and shows 
no correlation with RV (r=0.22) suggesting a Keplerian character of the observed RV variations.  

We therefore assume  that in both cases the observed RV variations are Keplerian,
at least in the first approximation, but we will come back to this issue in Section \ref{activity}.

\section{Keplerian analysis \label{results-g}}

\begin{figure}
   \centering
   \includegraphics[width=0.5\textwidth]{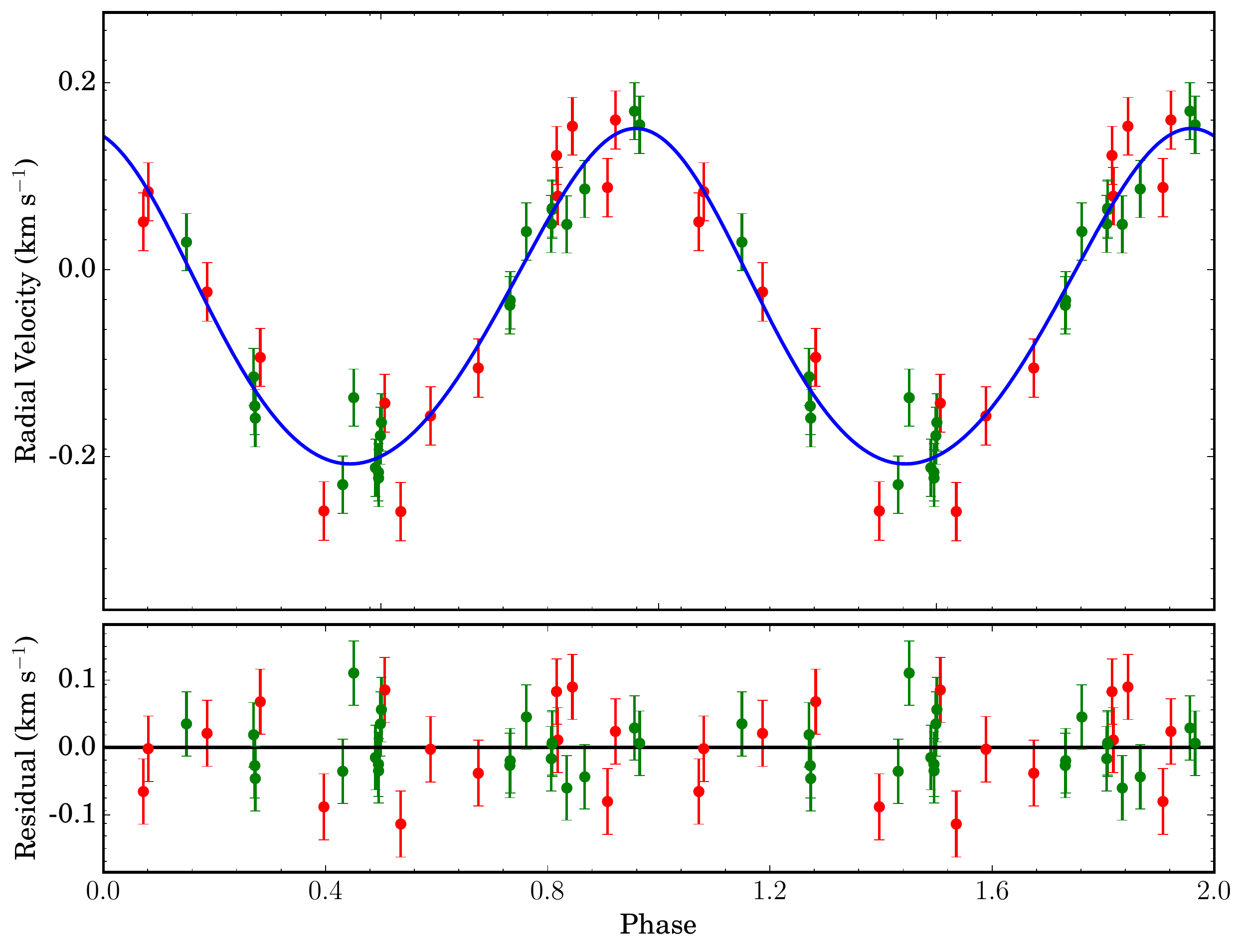}
   \caption{Keplerian orbital fit for HD 5583 b. In red HET HRS data, in green TNG HARPS-N data. 
   Jitter is added to RV uncertainties. }
   \label{Fit_1}
\end{figure} 

\begin{figure}
   \centering
   \includegraphics[width=0.5\textwidth]{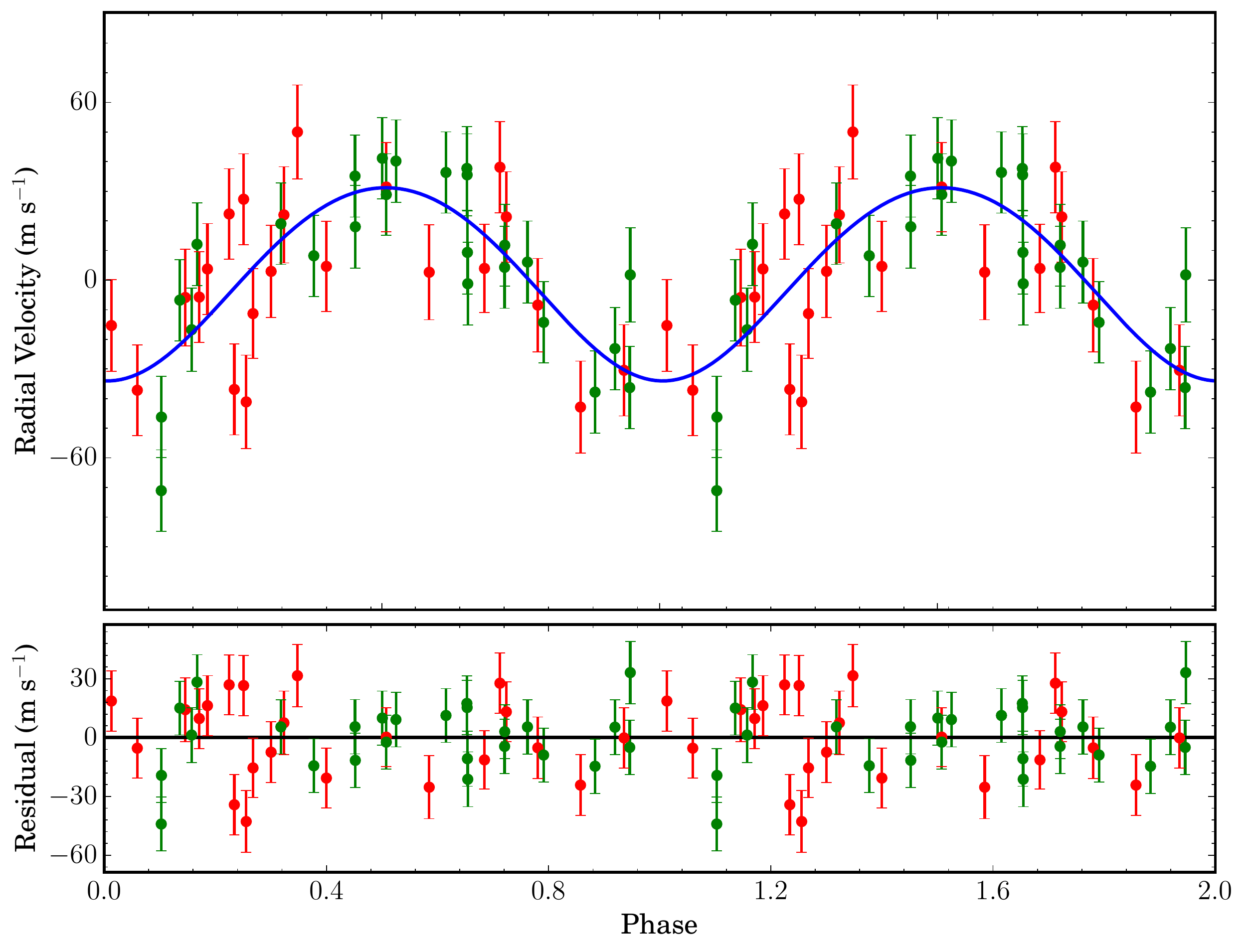}
   \caption{Keplerian orbital fit for BD+15 2735 b. In red HET HRS data, in green TNG HARPS-N data. 
   Jitter is added to RV uncertainties. }
   \label{Fit_2}
\end{figure} 

Keplerian orbital parameters  were derived using a hybrid approach 
(e.g. \citep{2003ApJ...594.1019G, 2006A&A...449.1219G, 2007ApJ...657..546G}), 
in which the PIKAIA-based, global genetic algorithm (GA; \cite{Charbonneau1995}) was combined with 
a faster and more precise local method, namely MPFit algorithm \citep{Markwardt2009},
to find the best-fit Keplerian orbit  delivered by RVLIN \citep{WrightHoward2009}
modified  to allow the stellar jitter to be fitted as a free parameter \citep{2007ASPC..371..189F, 2011ApJS..197...26J}.
The clearly visible  periodic signals found in the RV data with the Lomb-Scargle (LS) 
periodogram  \citep{1976ApJSS..39..447L, 1982ApJ...263..835S, 1992nrfa.book.....P} 
- Figure \ref{LSP1} and \ref{LSP2} were used as  first approximations of orbital periods.

The RV bootstrapping  method  \citep{1993ApJ...413..349M, 1997A&A...320..831K, Marcy2005,Wright2007}  
was employed to assess the uncertainties of the best-fit orbital parameters defined 
as the width of the resulting distribution of $10^6$ trials of scrambled data
between the 15.87th and 84.13th percentile. The false alarm probability (FAP) of the final 
orbital solutions was derived by repeating the whole hybrid  analysis on $10^5$ sets of scrambled data.

The results of Keplerian analysis are presented in Table \ref{KeplerianFits} and in Figures \ref{Fit_1} and \ref{Fit_2}.

We note that in the case of BD+15~2735 the resulting RV jitter is consistent with our estimates 
while  in the case of HD~5583 it is  $\approx4$ times larger than our estimate based 
on the empirical scaling relations of  \cite{KjeldsenBedding1995}.

\begin{table}
\centering
\caption{Keplerian orbital parameters of  HD~5583~b and  BD+15~2735~b.}
\input{table7}
\label{KeplerianFits}
\tablefoot{V$_{0}$ denotes the system velocity in the HET velocity system which is shifted off the HARPS-N velocity system by offset. The $\sigma_{jitter}$ is stellar intrinsic jitter as defined in \cite{2011ApJS..197...26J}.}
\end{table}

\section{Activity \label{activity}}

Giants are known to be variable stars (Payne-Gaposchkin 1954; Walker et al. 1989).
Moreover, both stars show Keplerian motion periods shorter than the maximum rotation periods
estimated from the projected rotation velocity and radii. Also in the case of BD+15~2735, 
for HET/HRS data a weak correlation between RV and BS exists. We  therefore need to discuss 
the available data in the context of stellar activity and its possible influence on the observed RV variations.

\subsection{Photometry}

HD~5583 was observed within Super WASP \citep{Pollacco2006} over 1243 days between JD 2453209.6 to 2454452.5,
partly contemporaneous with our  HET observations. The 4367 epochs of photometric 
data show an average brightness of $7.79 \pm 0.04$~mag and contain no detectable 
periodic signal - Figure \ref{LSP1}.  We note, however, that the uncertainty in the Super WASP
data is much larger than expected for such a  bright star and it is most likely 
of instrumental nature, the star is simply too bright. 

For BD+15~2375 two sets of photometric data are available:
367 epochs of ASAS \citep{1997AcA....47..467P} observations and 83 epochs of NSVS data \citep{2004AJ....127.2436W}.
The ASAS data were collected over 2400 days between HJD 2452624.9 and 2455025.5
and are partly contemporaneous with our HET data. These data show mean brightness of $10.294 \pm 0.018$~mag.
The NSVS data were collected over 336 days between JD 2451274.8 and 2451611.9 
i.e. long before our observations started. The data show mean brightness of $10.189 \pm 0.015$~mag. 
In both cases the uncertainties are consisted with the stellar brightness 
and the photometric data contain no detectable periodic signal - Figure \ref{LSP2}.

The available photometric data exclude pulsations as a possible source
of observed RV variations of our program stars. The lack of observable periodic photometric variations
also excludes the stellar spot scenario to explain the RV variations. We may, however, in a very conservative approach, 
 use the estimated uncertainty in the available photometric data sets as a proxy 
 of the spot size and estimate what  would be the maximum contribution due to a hypothetical spot on the RV and BS signal.
 
Following the approach of \cite{Hatzes2002} we may estimate that in the case of HD~5583
the maximum peak-to-peak RV variations due to such a hypothetical spot might reach $225 \ms$ 
and the BS - $36 \ms$. The observed RV signal in both HET HRS and TNG HARPS-N data
is over twice that much, hence a spot interpretation is unlikely.  
 
In the case of BD+15~2735 the maximum RV variations due to a spot might 
reach $126 \ms$ and  BS - $17 \ms$ according to \cite{Hatzes2002}.
These values are comparable to the observed ones.

\subsection{Spectral line bisectors}

The HET and HRS and HARPS-N BS are not directly comparable because they 
were calculated not only from different instruments but also from different  sets of spectral lines. 
{ The HET bisecor span was calculated from a cross-correlation function constructed from spectra cleaned from the I2 lines and correlated with a syntetic K2 spectrum. It was defined as a difference between CCF profile mean velocities at 0.1-0.25 and  0.65-0.80 of the line depth  \citep{Nowak2012}. The HARPS-N CCF line bisectors were calculated  
as mean velocity difference between  0.1-0.4  and 0.55-0.9 of the line depth \citep{2001A&A...379..279Q}.
} Thus, we need to  consider them separately. 

For HD~5583 HET/HRS BS shows an amplitude of $100 \ms$ with average uncertainty of $19 \ms$, 
while for HARPS-N the amplitude is $41 \ms$. These variations are comparable, or larger than those 
due to a hypothetical spot. They are, however, 5-12 times smaller than the observed RV variations 
and show no correlation with RV. Any significant contribution to RV from BS variations is therefore unlikely.

HET/HRS BS data for BD+15~2375 show an amplitude of $99 \ms$ and average uncertainty of $17 \ms$. 
The HARPS-N BS vary by $40 \ms$. These variations are larger than those produced 
by a hypothetical spot. The amplitude of BS variations is  comparable (HET data) 
or only $\approx2$ times lower (TNG) than the observed RV variations.
 
We note that HET RVs show larger amplitude than that of HARPS-N. Also HET BSs,
although not directly comparable to HARPS-N ones, show $\approx3$ times larger variations.
The amplitude of BS variations is roughly the same as in RV.  Also intriguing is 
the RV - BS correlation in these data.  Given the relatively low number of observations 
it may very well prove spurious, especially that the most recent HARPS-N data show 
lower RV and BS amplitudes and no RV-BS correlation.

It cannot be excluded then that the HET data were gathered during a period of increased activity 
of BD+15~2735. A period of enlarged spot moving with the rotating star,  that resulted
in the increased RV signal and RV-BS correlation.
HARPS-N data, showing a  periodic RV signal  and no RV-BS correlation, 
proves instead that only a fraction, if any at all, of HET RV
was due to activity and that the Keplerian signal dominates.
 The period
of the activity cycle is not known but the proposed scenario seems to be additionally supported by existing
photometry data contemporaneous to our HET observations, that show larger uncertainties than older NSVS data.
This is somewhat similar to the periodic activity detected in GJ 581 as discussed in \cite{2014Sci...345..440R}.
 
\subsection{Ca H\&K lines}

The Ca II H and K line profiles are widely accepted as stellar activity indicators 
(e.g., \citealt{ Noyes1984, Duncan1991}) and if the reversal profile, typical for active stars
\citep{EberhardSchwarzschild1913} is present,  chromospheric activity can be deduced.

Unfortunately, neither the Ca~II H and K lines nor the infrared Ca~II triplet lines at
849.8-854.2~nm are available in HET and HRS spectra, but the Ca~II H
and K lines are present in the TNG HARPS-N spectra. The S/N ratio of our red giants in that spectral 
range is low but we found no trace of  reversal. To quantify the observations we
calculated an instrumental $\shk$ index according to the prescription of \cite{Duncan1991} 
for  HARPS-N data. For HD 5583 we obtained a value of 0.17 with a standard deviation 
of 0.08, a typical value  for non-active  stars. The $\shk$ for this star shows no 
statistically significant correlation with the RV (r=-0.22). 
 
Also in the case of BD+15~2735 no trace of profile reversal was found,
$\shk=0.19 \pm 0.03$ and there exists no correlation with the RV (r=-0.01).
 
We conclude thet Ca II H and K line profiles analysis thus reveals  that over the period
covered by TNG observations both giants are rather non-active and there is no trace 
of influence of activity upon the observed RV variations.

\subsection{H$_{\alpha}$}

Another activity indicator that we can use for the HET and HRS spectra
is the H$_\alpha$ (656.2808 nm) line, which in red giants is rather weak. 
We measured the H$\alpha$ index ($I_{\mathrm{H}\alpha}$) following the procedure described in detail in
\cite{2013AJ....146..147M} based on the approach presented by
\cite{2012A&A...541A...9G} and \citet[][and references therein]{2013ApJ...764....3R}.  We analyze the H$\alpha$ index measurements from HET and TNG spectra instruments together applying the RV offset determined in Keplerian anlysis.
 


In the case of HD 5583 the $I_{\mathrm{H}\alpha}$  has a mean value of 0.034 and varies by 0.003, or $10.0\%$.
These variations are not correlated with the RV (r=0.12). The control line (FeI 6593.878 $\AA$) index 
has a mean value of 0.012,  varies by 0.001 or $9.0\%$ and is not correlated with RV (r=-0.13). The variations in both indexes 
are correlated (r=0.55) which proves that  the observed  H$_{\alpha}$ line profile variations 
are mostly due to instrumental effects and do not carry information on stellar variability in H$_\alpha$  line.



 BD+15 2735  has a mean $I_{\mathrm{H}\alpha}$  value of 0.035 and varies by 0.004
or $12.6\%$. It shows no correlation with RV (r=-0.23). At the same time the control line 
index has a mean value of 0.011, varies by 0.001 or $10.3\%$ and  shows no correlation with RV (r=-0.14).
These variations again show correlation (r=0.52) what 
proves instrumental nature of the observed H$_{\alpha}$ line profile variations. 

We may conclude therefore that the analysis of $I_{\mathrm{H}\alpha}$ index based 
on HET/HRS data excludes significant level of activity in both stars.

\begin{figure}
   \centering
   \includegraphics{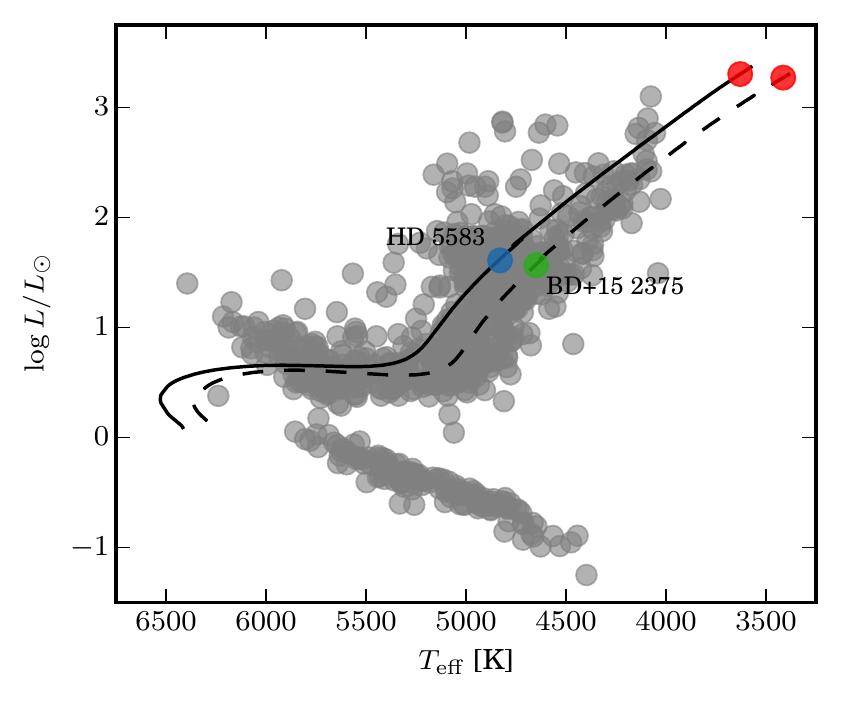}
   \caption{Hertzsprung-Russell diagram for the complete PTPS sample 
   with HD~5583 and BD+15~2735 and their evolutionary tracks. In red we show the stage when stellar radius reaches current planetary orbits. }
   \label{HRD}
\end{figure} 

\section{Discussion  \label{conclusions}}

We presented two giants with RV variations that, in the face of lack of compelling evidence 
of them being caused by stellar activity, we interpret as Keplerian motion due to low-mass companions.
The giant star HD 5583 has a 
minimum mass $5.78\MJ$ companion 
in a 0.529~AU,
nearly circular ($e=0.076$) orbit, 
while the star BD+15~2735 has a 
minimum mass $1.061\MJ$ 
companion in a 0.576~AU, 
nearly circular ($e=0.01$) orbit. 
Assuming an average value of $\sin i$ both 
these companions stay within the planetary-mass range.

HD~5583~b has a mass of a rather typical 
gas giant planet but its nearly circular a=0.529 au orbit 
is the tightest  around solar-type ($M<1.2 M_{\odot}$) evolved stars
 detected with the RV technique so far. It can be only compared to 
 our own BD+15~2940~b \citep{Nowak2013} at 0.539~AU
 or HD 32518 b at 0.59~AU \citep{2009A&A...505.1311D}.

BD+15~2735~b is a very rare Jupiter-mass companion
to an evolved star. Actually it is the lightest planet detected with the RV technique. 
Only two other similar  planets are known, both from PTPS:
$m_{\mathrm{p}}\sin i =1.16 \MJ$ BD +48 738b \citep{Gettel2012a,Niedzielski2015c}
and $m_{\mathrm{p}}\sin i=1.17 \MJ$  BD+15 2940 b \citep{Nowak2013, Niedzielski2015c}.

In Figure \ref{HRD} evolutionary tracks for both stars are given and, if their inferred location
in the HR diagram is correct, both stars are evolving up the RGB. From their stellar parameters
we estimate that their current orbits will lead to planet engulfment in  
82 and 89 mln years,
respectively \citep{VillaverLivio2009,Villaver2014}. This is the expected fate of most planets known
so far. 

 Both planets are within 0.6 au and thus orbit
 at the very edge of the planet-avoidance zone around evolved stars.

\begin{figure}
   \centering
   \includegraphics[width=0.5\textwidth]{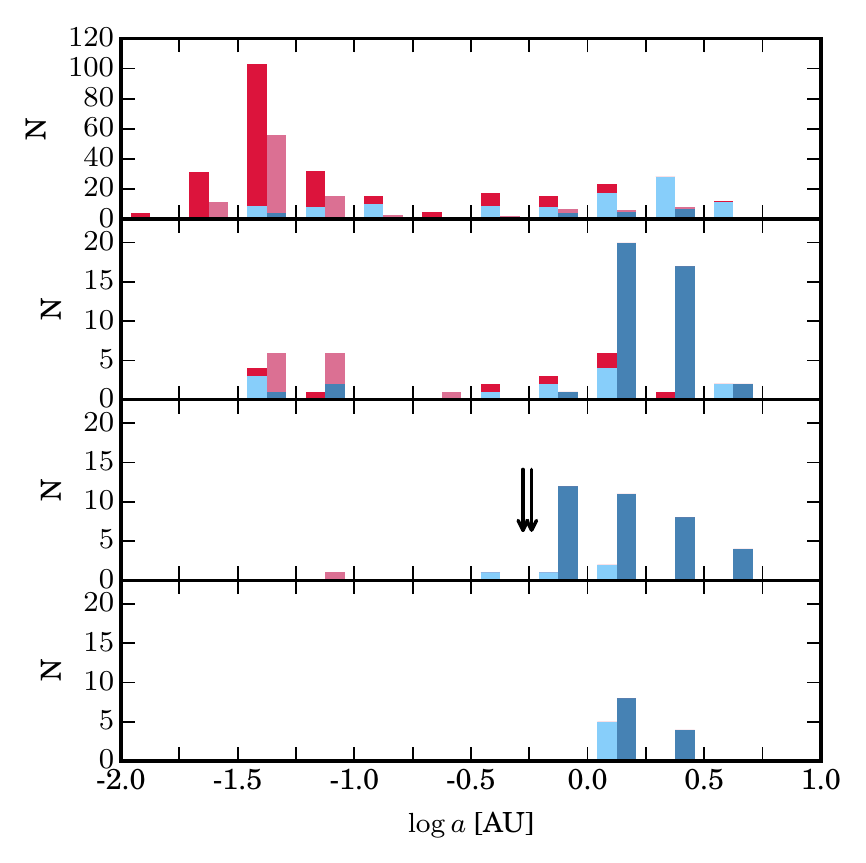}
   \caption{Semi-major axis distributions for (top to bottom) dwarfs (here $\log g=4.5\pm0.5$),
   subgiants (here $\log g=3.5\pm0.5$), giants (here $\log g=2.5\pm0.5$) 
   and bright giants (here $\log g=1.5\pm0.5$). In blue - RV planets only, in red all other technique planets including Kepler.
   In every 0.25 dex range two bars are presented, left  for hosts 
   of  $\leq1.2 M/M_{\odot}$,  and right  $> 1.2 M/M_{\odot}$.
   Planets with very extended orbits are omitted for clarity. The positions of the two planets presented in this paper are indicated by arrows.}
   \label{histogram}
\end{figure}

In order to put the newly discovered planets in a perspective
we show in Fig. \ref{histogram} the orbital distribution of the known planets separated by
the evolutionary status of the star using as proxy the stellar $\log g$ value. 
This figure illustrates well how efficient our project is in detecting planetary-mass 
companions to evolved stars prone to tidal interactions with their hosts in the near future.

We note, in passing, that the RV jitter of HD~5583  is 3-4 times higher
than expected for a star of this mass and luminosity.
Its mass, although uncertain as estimated by means of Bayesian analysis,
is not expected to be much different from the current estimate.  It is not excluded then,
that the luminosity of this object is underestimated by such a factor and that the star 
is actually a horizontal branch object already. If that were the case, this
planet might have survived the RGB evolution of its host either surviving engulfment on the envelope or having its orbital distance decreased through the mechanism identified in \cite{VillaverLivio2009}.
Then HD~5583 b would be one of very few companions 
to survive evolution of its host  \citep{Niedzielski2015b}.

\begin{acknowledgements}
 
 We thank the HET  and TNG  resident astronomers and telescope operators  for support.

AN, MoA, GN, BD, and MiA  were supported by the Polish National Science Centre grant UMO-2012/07/B/ST9/04415.
MoA also acknowledges the Mobility+III fellowship from the Polish Ministry of Science and Higher Education.
KK was funded in part by the Gordon and Betty Moore Foundation's
Data-Driven Discovery Initiative through Grant GBMF4561.
This research was supported in part by PL-Grid Infrastructure.
EV work was supported by the Spanish Ministerio de Ciencia e Innovacion (MICINN),
Plan Nacional de Astronomia y Astrofisica, under grant AYA2013-45347P. 
AW was supported by the NASA grant NNX09AB36G. 

The HET is a joint project of the University of Texas at Austin, the Pennsylvania State University, Stanford University, Ludwig-
Maximilians-Universit\"at M\"unchen, and Georg-August-Universit\"at G\"ottingen.
The HET is named in honor of its principal benefactors, William P. Hobby and Robert E. Eberly.

The Center for Exoplanets and Habitable Worlds is supported by the Pennsylvania State University,
the Eberly College of Science, and the Pennsylvania Space Grant Consortium. 

This research has made use of the SIMBAD database, operated at CDS (Strasbourg, France) and NASA's Astrophysics Data System Bibliographic Services.

This research has made use of the Exoplanet Orbit Database and the Exoplanet Data Explorer at exoplanets.org.

\end{acknowledgements}

\bibliographystyle{aa} 
\bibliography{an.bib} 
\end{document}

%% file: table7.tex
{\tabulinesep=0.4mm
\begin{tabu}{lll}

\hline
Parameter & HD 5583 b & BD+15 2735 b  \\
\hline
\hline

$P$ (days)                                        & $139.35^{+0.21}_{-0.22}$  & $153.22^{+0.44}_{-0.44}$ \\
$T_0$ (MJD)                                       & $56021^{+18}_{-17}$       & $57680^{+700}_{-630}$ \\
$K$ ($\textrm{m s}^{-1}$)                         & $225.8^{+4.5}_{-4.0}$     & $38.3^{+2.3}_{-1.0}$ \\
$e$                                               & $0.076^{+0.070}_{-0.023}$ & $0.001^{+0.25}_{-0.001}$ \\
$\omega$ (deg)                                    & $12^{+330}_{-6}$          & $55^{+240}_{-17}$ \\
$m_2\sin i$ ($\textrm{M}_{\textrm{J}}$)           & $5.78 \pm 0.53$           & $1.061 \pm 0.27$ \\
$a$ (AU)                                          & $0.53 \pm 0.02$           & $0.576 \pm 0.027$ \\
$V_0$ ($\textrm{m s}^{-1}$)                       & $-37.8^{+7.5}_{-7.1}$     & $1.2^{+1.8}_{-1.7}$ \\
offset ($\textrm{m s}^{-1}$)                      & $12060^{+21}_{-21}$       & $-9322^{+7}_{-8}$\\
$\sigma_{\textrm{jitter}}$ ($\textrm{m s}^{-1}$)  & $38.2^{+2.0}_{-3.2}$      & $13.7^{+2.0}_{-0.3}$\\
$\sqrt{\chi_\nu^2}$                               & $1.88$                    & $1.6$\\
$\sigma_{\textrm{RV}}$ ($\textrm{m s}^{-1}$)      & $69.5$                    & $22.82$\\
$N_{\textrm{obs}}$                                & $36$                      & $48$ \\

\hline

\end{tabu}
}